\documentclass{aa}

\usepackage[varg]{txfonts}
\usepackage{sidecap}
\usepackage{dblfloatfix}
\usepackage{color}

\usepackage{hyperref}
\hypersetup{
 colorlinks,
 citecolor=blue,
 linkcolor=red,
 urlcolor=blue}
 
\begin{document}

\title{Heating of the solar chromosphere in a sunspot light bridge by electric currents}

\titlerunning{Heating of the solar chromosphere in a sunspot by electric currents}

\authorrunning{Louis, Prasad, Beck, Choudhary, \& Yalim}

\author{Rohan E. Louis\thanks{Corresponding author: rlouis@prl.res.in}\inst{1}
\and Avijeet Prasad\inst{2}
\and Christian Beck\inst{3}
\and Debi P. Choudhary\inst{4}
\and Mehmet S. Yalim\inst{2}
}

\institute{Udaipur Solar Observatory, Physical Research Laboratory,
Dewali Badi Road, Udaipur - 313001, Rajasthan, India
\and Center for Space Plasma and Aeronomic Research,
The University of Alabama in Huntsville, Huntsville AL 35899, USA
\and National Solar Observatory (NSO), 
3665 Discovery Drive, Boulder, CO 80303, USA
\and Department of Physics and Astronomy, 
California State University, Northridge (CSUN), CA 91330-8268, USA
}

\date{Received 02 June 2021 / Accepted 13 July 2021}

\abstract {Resistive Ohmic dissipation has been suggested as a mechanism for heating the solar chromosphere, but few 
studies have established this association.} 
{We aim to determine how Ohmic dissipation by electric currents can heat the solar chromosphere.} 
{We combine high-resolution spectroscopic \ion{Ca}{II} data from the Dunn Solar 
Telescope and vector magnetic field observations from the Helioseismic and Magnetic Imager (HMI) to investigate 
thermal enhancements in a sunspot light bridge. The photospheric 
magnetic field from HMI was extrapolated to the corona using a non-force-free field technique that provided the 
three-dimensional distribution of electric currents, while an inversion of the chromospheric \ion{Ca}{II} line with 
a local thermodynamic equilibrium and a nonlocal thermodynamic equilibrium spectral archive delivered the temperature 
stratifications from the photosphere to the chromosphere.} 
{We find that the light bridge is a site of strong electric currents, of about 0.3\,A\,m$^{-2}$ at the bottom boundary,
which extend to about 0.7\,Mm while decreasing monotonically with height. These currents produce a chromospheric 
temperature excess of about 600--800\,K relative to the umbra. Only the light bridge, where relatively weak and 
highly inclined magnetic fields emerge over a duration of 13\,hr, shows a spatial coincidence of thermal 
enhancements and electric currents. The temperature enhancements and the Cowling heating are primarily 
confined to a height range of 0.4--0.7\,Mm above the light bridge. The corresponding increase in internal energy 
of 200\,J\,m$^{-3}$ can be supplied by the heating in about 10\,min.} 
{Our results provide direct evidence for currents heating the lower solar chromosphere through Ohmic dissipation.} 


\maketitle

\section{Introduction}
\label{intro}
The solar chromosphere serves as an important conduit for mass and energy 
between the dense, 6000\,K photosphere and the tenuous, million degree corona. 
The solar chromosphere has a complex magnetic structure, where the plasma 
beta changes dramatically \citep{2001SoPh..203...71G}.
Determining the processes that maintain the thermal structure of the solar 
atmosphere is one of the fundamental problems in solar physics \citep{1996SSRv...75..453N}.

\begin{figure*}[!ht]
\centerline{
\includegraphics[angle=0,width=0.314\textwidth]{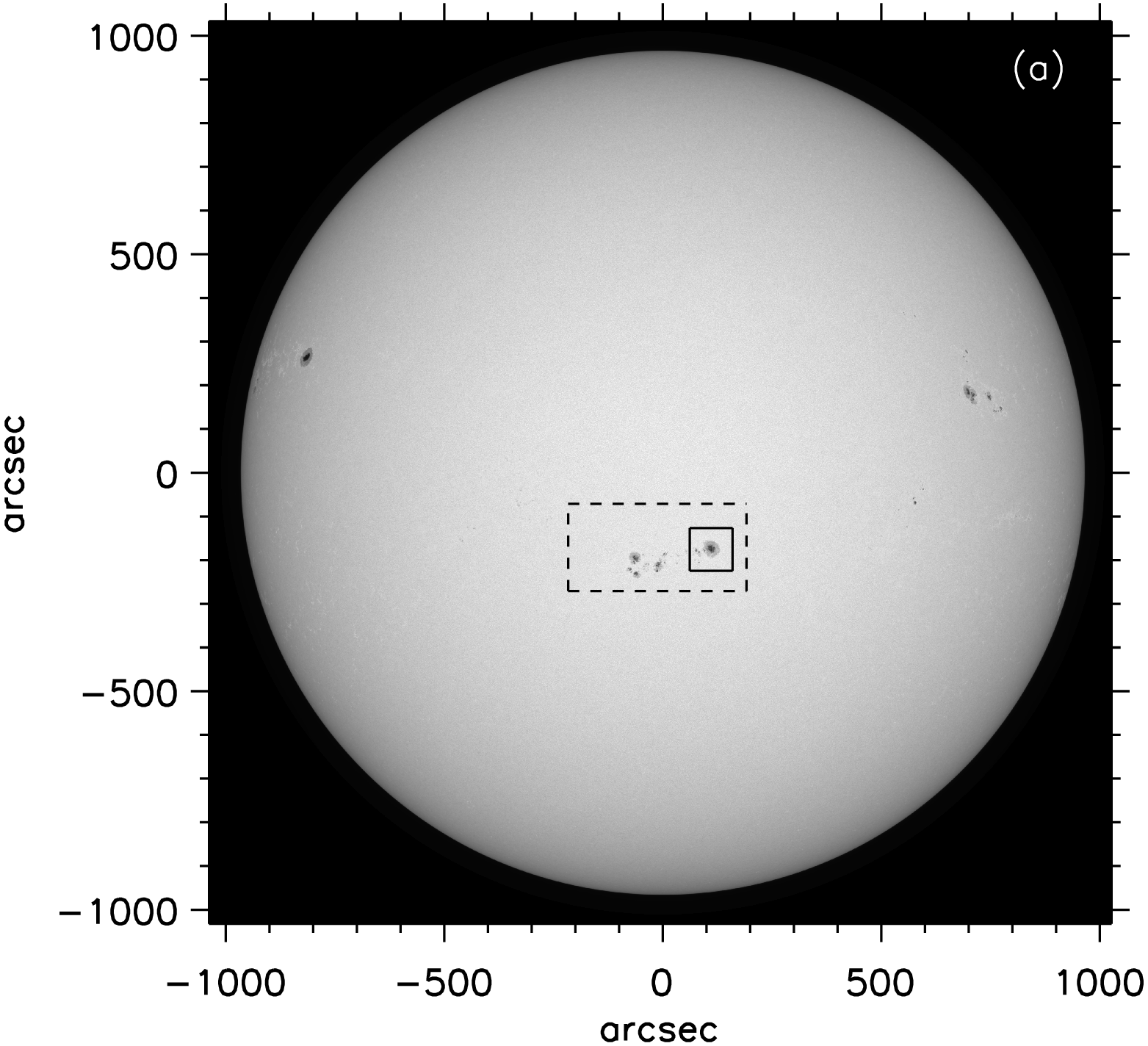}
\hspace{-5pt}
\includegraphics[angle=90,width=0.6\textwidth]{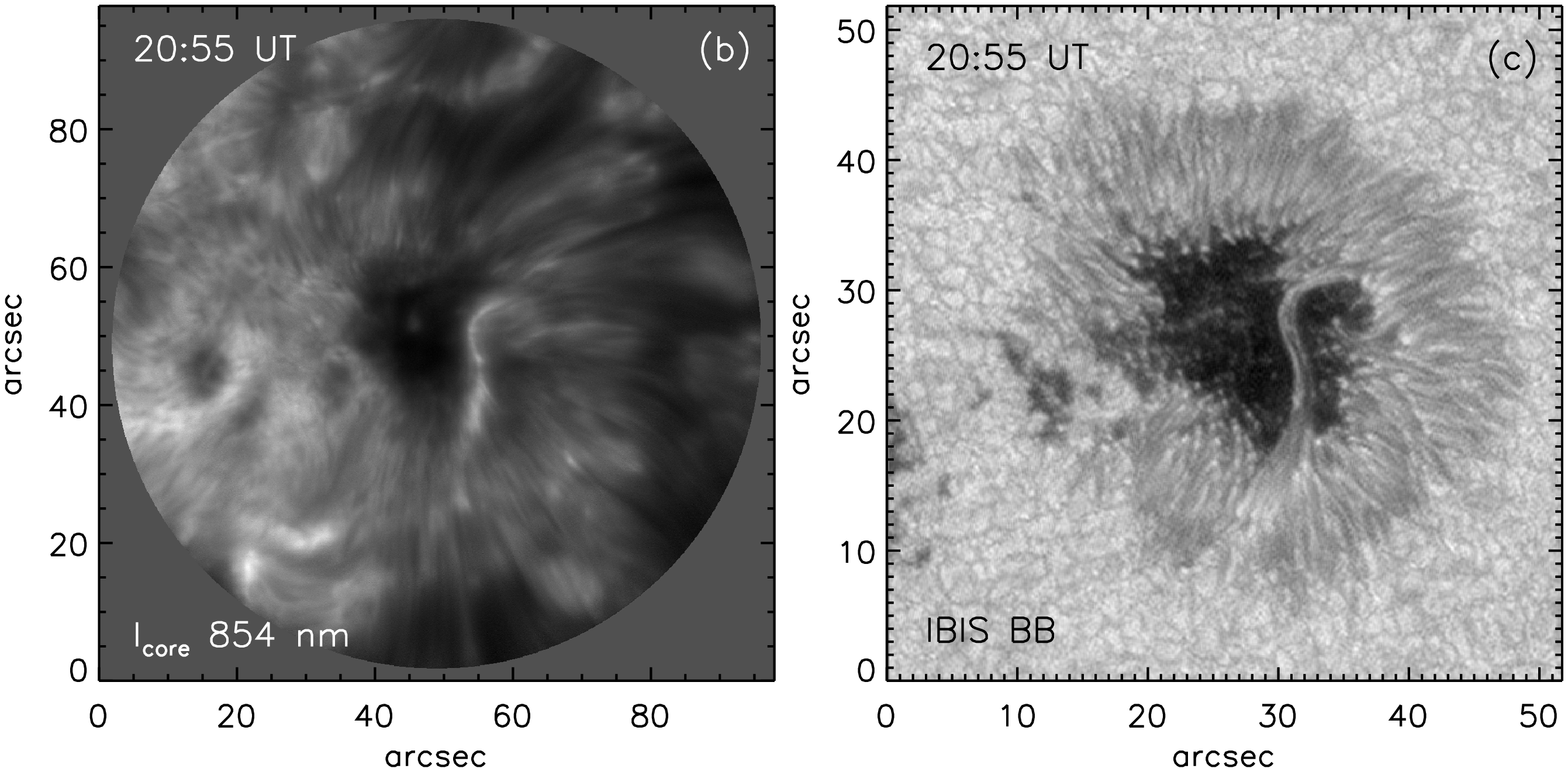}
}
\vspace{-225pt}
\centerline{
\hspace{60pt}
\includegraphics[angle=90,width=1.1\textwidth]{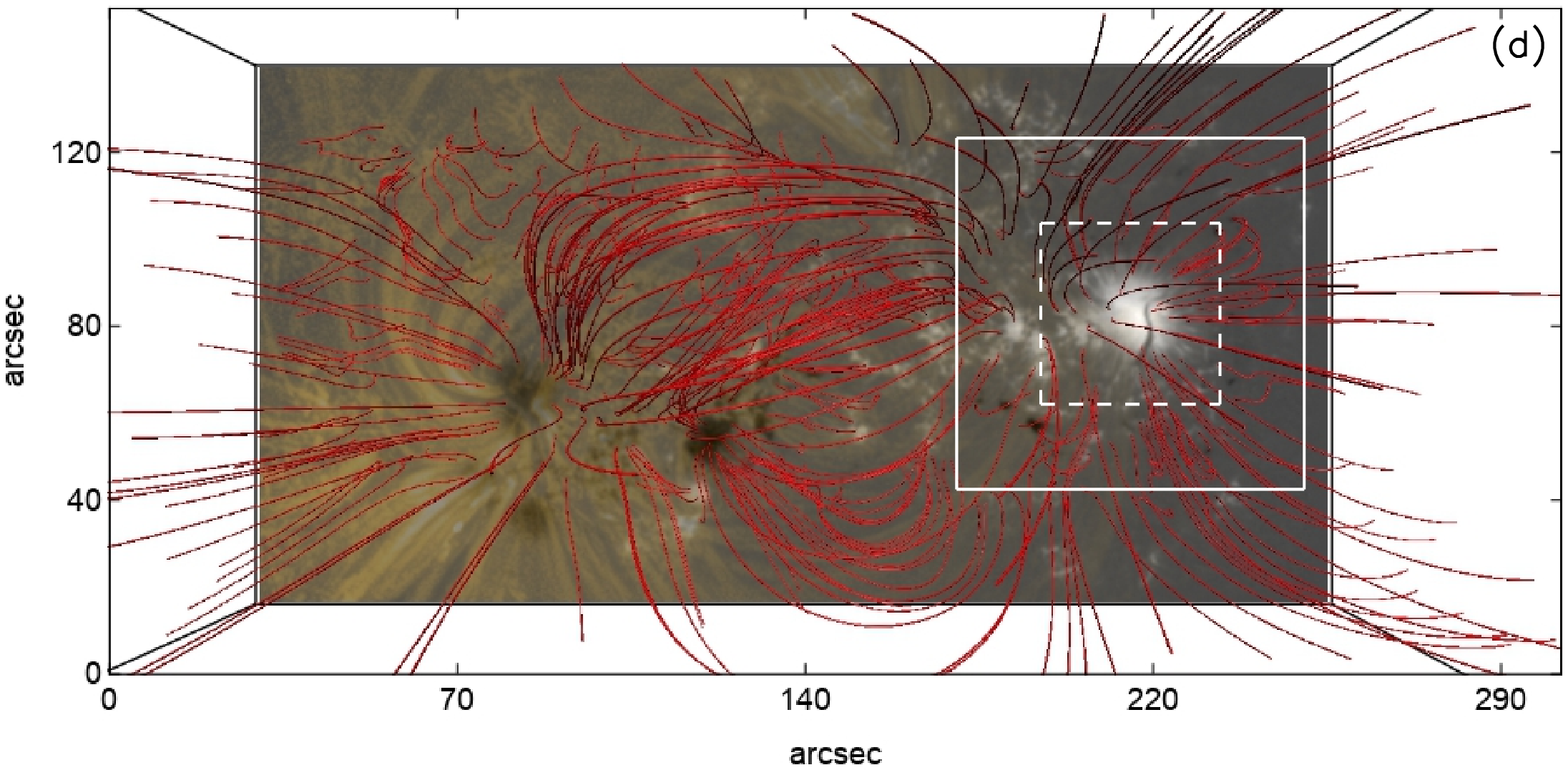}
}
\vspace{-20pt}
\caption{HMI and IBIS observations of the leading sunspot in NOAA AR 12002.
Top: Full-disk HMI continuum image (left) at 21:00 UT on 2014 March 13, 
IBIS \ion{Ca}{II} IR line-core image (middle), and speckle-reconstructed IBIS 
broadband image (right). The square and dashed rectangle in panel a 
indicate the IBIS and HMI SHARP FOV, respectively. Bottom: Field lines derived 
from the NFFF extrapolation overlaid on a composite image of the vertical 
component of the magnetic field and the AIA 171~\AA\ image for the SHARP FOV. 
The solid and dashed white squares correspond to the IBIS FOV and the smaller 
FOV shown in Fig.~\ref{fig02}, respectively.} 
\label{fig01}
\end{figure*}

The energy transfer in the chromosphere can be attributed to a number of mechanisms, such as Alf\'ven waves 
\citep{1961ApJ...134..347O,1981ApJ...246..966S,2011ApJ...736....3V,2016ApJ...819L..11S,2018NatPh..14..480G, 2020ApJ...900..120S}, 
spicules \citep{1968SoPh....3..367B,1978SoPh...57...49P,1982ApJ...255..743A,2009ApJ...701L...1D,2016SoPh..291.2281B}, 
nanoflares \citep{2018ApJ...862L..24P,2020ApJ...891...52S}, and magneto-acoustic shocks \citep{2015ApJ...799L..12D}.  
Some authors have suggested that the heating of the chromosphere is due to the dissipation of acoustic waves 
\citep{1978A&A....70..487U,2007ApJ...671.2154K}, although this has been questioned by  
\cite{1982ApJ...255..743A} and \cite{2009A&A...507..453B,2012A&A...544A..46B}. 

Another candidate for heating the solar chromosphere is resistive Ohmic dissipation  
\citep{1983ApJ...264..635P,2008ApJ...686L..45T}. The occurrence of dynamic phenomena in the 
chromosphere and transition region has been attributed to plasma heating by the formation
of current sheets when a discontinuity in the three-dimensional (3D) magnetic field arises
\citep{2003Natur.425..692S,2021NatAs...5..237B}. Such currents are often seen in sunspot light 
bridges (LBs) and $\delta$ spots 
\citep{2006A&A...453.1079J,2009ApJ...696L..66S,2014A&A...562L...6B,2015ApJ...811..137T,2018A&A...609A..14R}. 
However, estimates of the current density are typically confined to the solar photosphere and  
only provide the vertical component of the current $J_z$, while \cite{2005ApJ...633L..57S} found 
only a weak correlation between currents and chromospheric heating.
In this article, we combine non-force-free field (NFFF) extrapolations with thermal 
inversions to study intense electric currents and strong chromospheric temperature 
enhancements over a sunspot, which occur as a result of large-scale magnetic flux 
emergence in an LB.

\begin{SCfigure*}
\begin{minipage}{0.92\textwidth}
\vspace{-10pt}
\centerline{
\hspace{-40pt}
\includegraphics[angle=0,width = 0.5\textwidth]{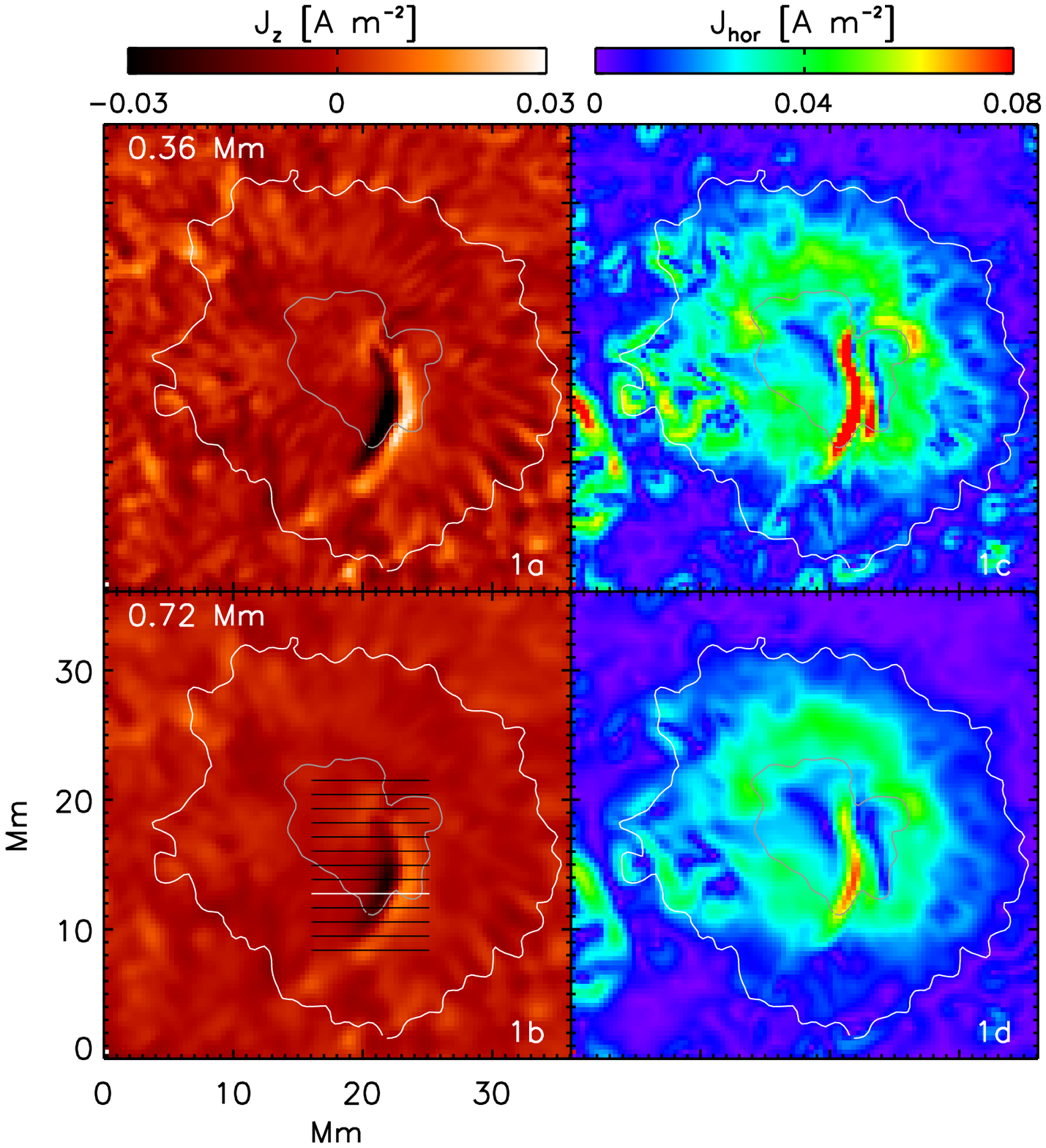} 
\hspace{-55pt}
\includegraphics[angle=0,width = 0.5\textwidth]{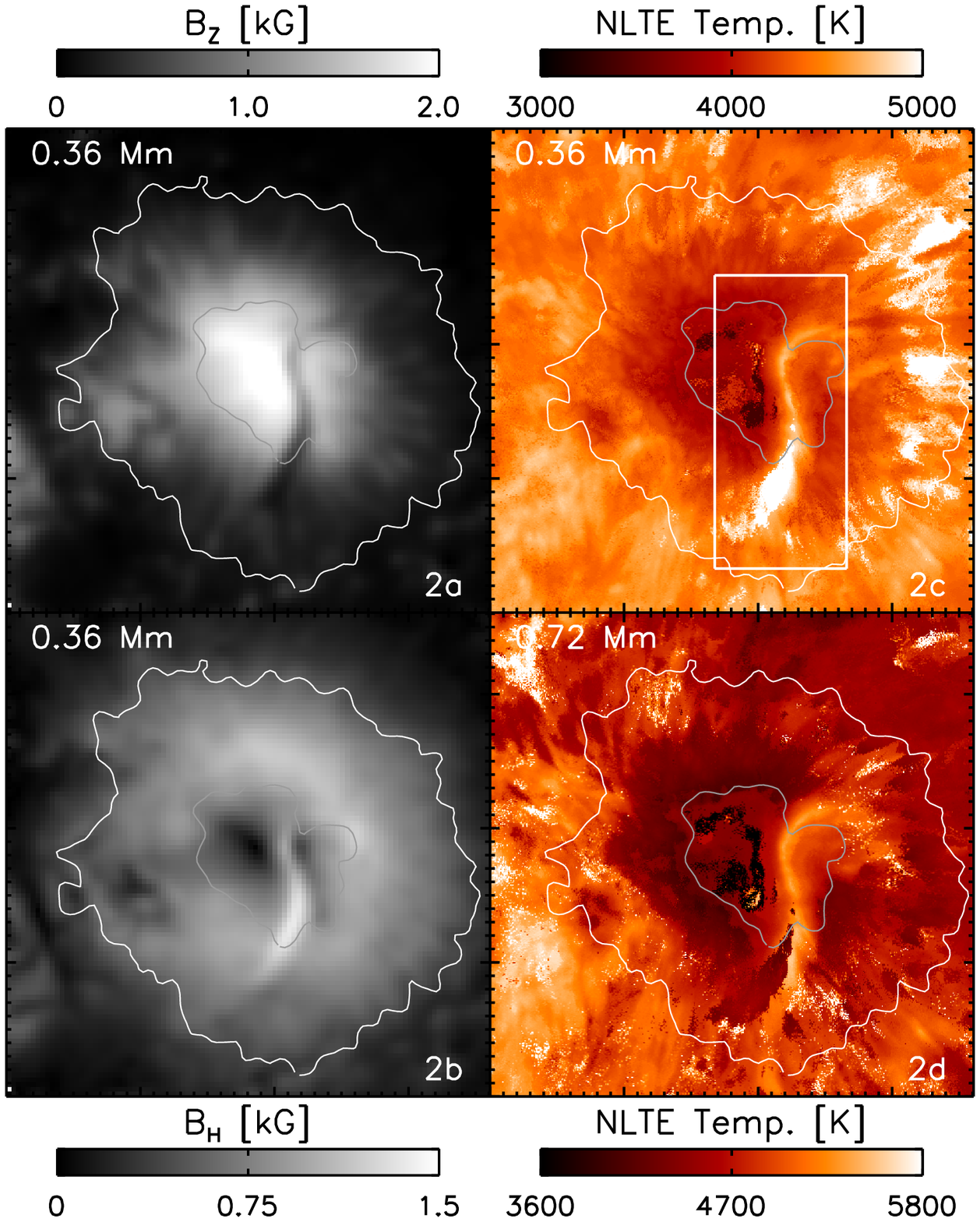}
}
\vspace{-88pt}
\centerline{
\hspace{-40pt}
\includegraphics[angle=0,width = 0.5\textwidth]{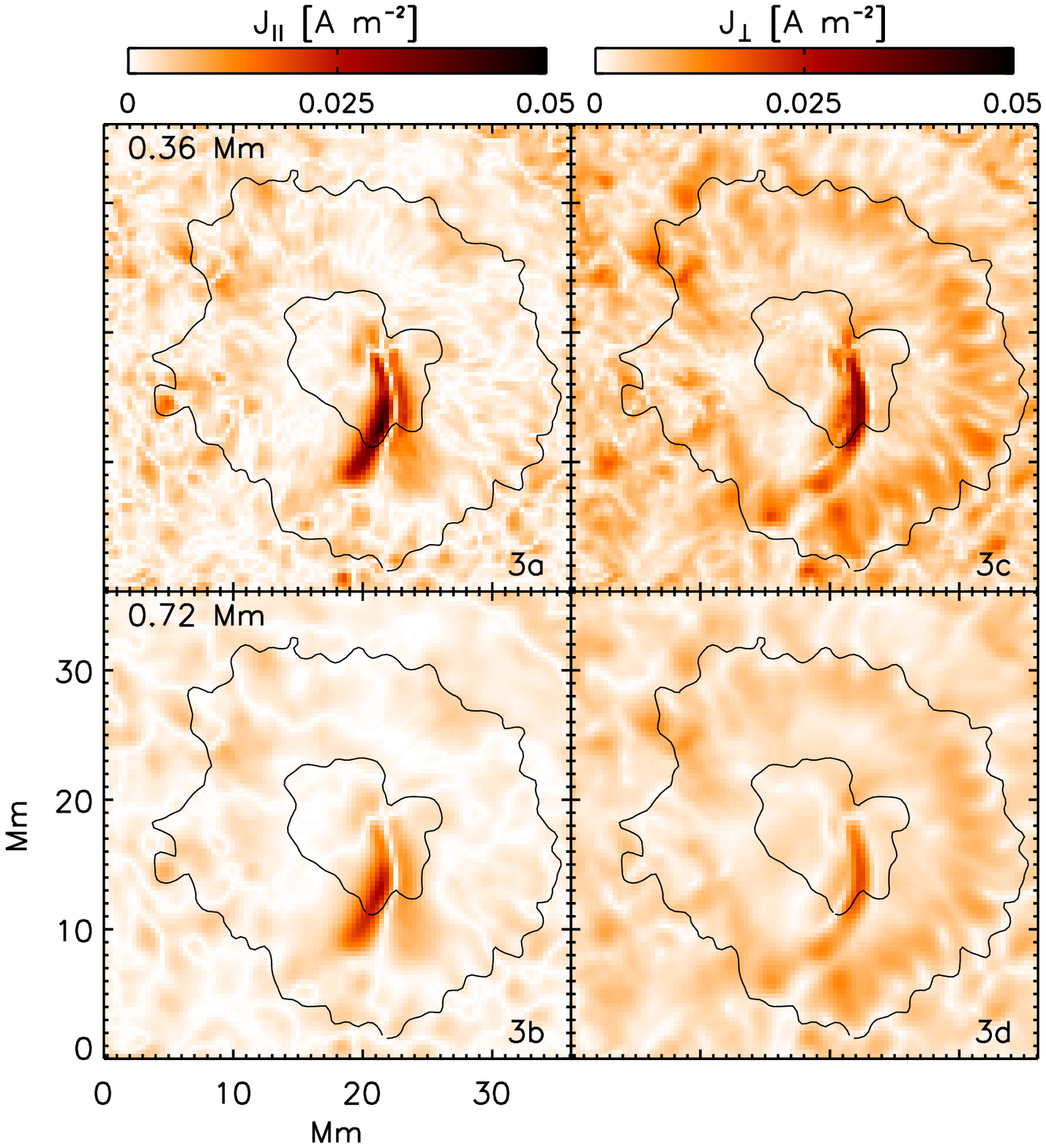} 
\hspace{-55pt}
\includegraphics[angle=0,width = 0.5\textwidth]{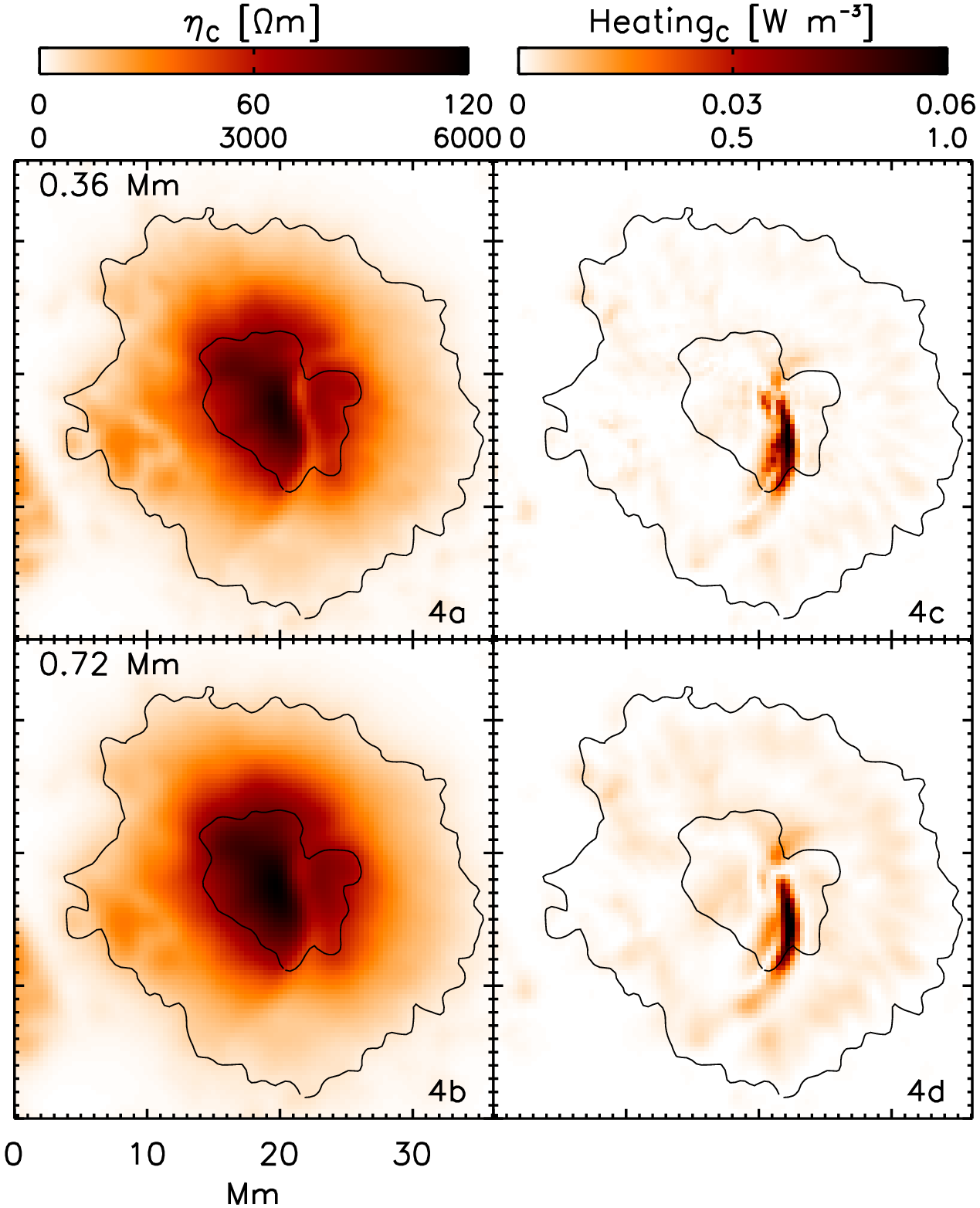}
}
\end{minipage}
\hspace{-75pt}
\vspace{-120pt}
\caption{Electric currents, temperature enhancements, and Joule heating in the sunspot LB. 
Panels 1a--1d: Spatial distribution of the vertical current density $J_z$ and 
horizontal current density $J_{hor}$ at z=0.36\,Mm (top) and z=0.72\,Mm (bottom). The 
horizontal lines in panel 1b indicate the locations of the 2D cuts shown in 
Fig.~\ref{fig04}. The white horizontal line marks cut No. 5 used in Figs.~\ref{fig04}g 
and \ref{fig04}h. Panels 2a--2b: Spatial distribution of the vertical (top) and horizontal 
(bottom) component of the magnetic field at z=0.36\,Mm. Panels 2c--2d: Spatial 
distribution of the NLTE temperature at z=0.36\,Mm (top) and z=0.72\,Mm (bottom). 
The white rectangle in panel 2c is a smaller FOV centered on the LB shown in 
Fig.~\ref{fig03}. Panels 3a--3d and 4a--4d: Spatial distribution of  
$J_{||}$, $J_{\bot}$, Cowling resistivity $\eta_{\textrm{\tiny{C}}}$, and its associated 
Joule heating. The color bar in panels 4 comprises two scales, where the top and bottom 
numbers correspond to 0.36\,Mm and 0.72\,Mm, respectively.}
\label{fig02}
\end{SCfigure*}

\section{Observations}
\label{obs}
We analyzed observations of the leading sunspot in NOAA AR 12002 (Fig.~\ref{fig01}a) 
on 2014 March 13 from 20:44 to 21:00 UT, combining data from the Helioseismic and Magnetic 
Imager \citep[HMI;][]{2012SoPh..275..229S} on board the Solar Dynamics Observatory 
\citep[SDO;][]{2012SoPh..275....3P} and the Interferometric BI-dimensional Spectrometer 
\citep[IBIS;][]{cavallini2006} at the Dunn Solar Telescope (DST).  The SDO data 
comprise two Spaceweather HMI Active Region Patch (SHARP) maps of the vector magnetic 
field at 20:48 UT and 21:00 UT. IBIS acquired spectral scans in the \ion{Ca}{II} IR line 
at 854\,nm with a spatial sampling of 0\farcs1 px$^{-1}$ across a 90$^{\prime\prime}$ 
circular field of view (FOV). Only the two scans at 20:44 UT and 20:55 UT were used here. 
The DST setup and data are described in detail in \citet{2020ApJ...905..153L}, hereafter LBC20.

\begin{figure}
\centerline{
\hspace{150pt}
\includegraphics[angle=0,width = 1.15\columnwidth]{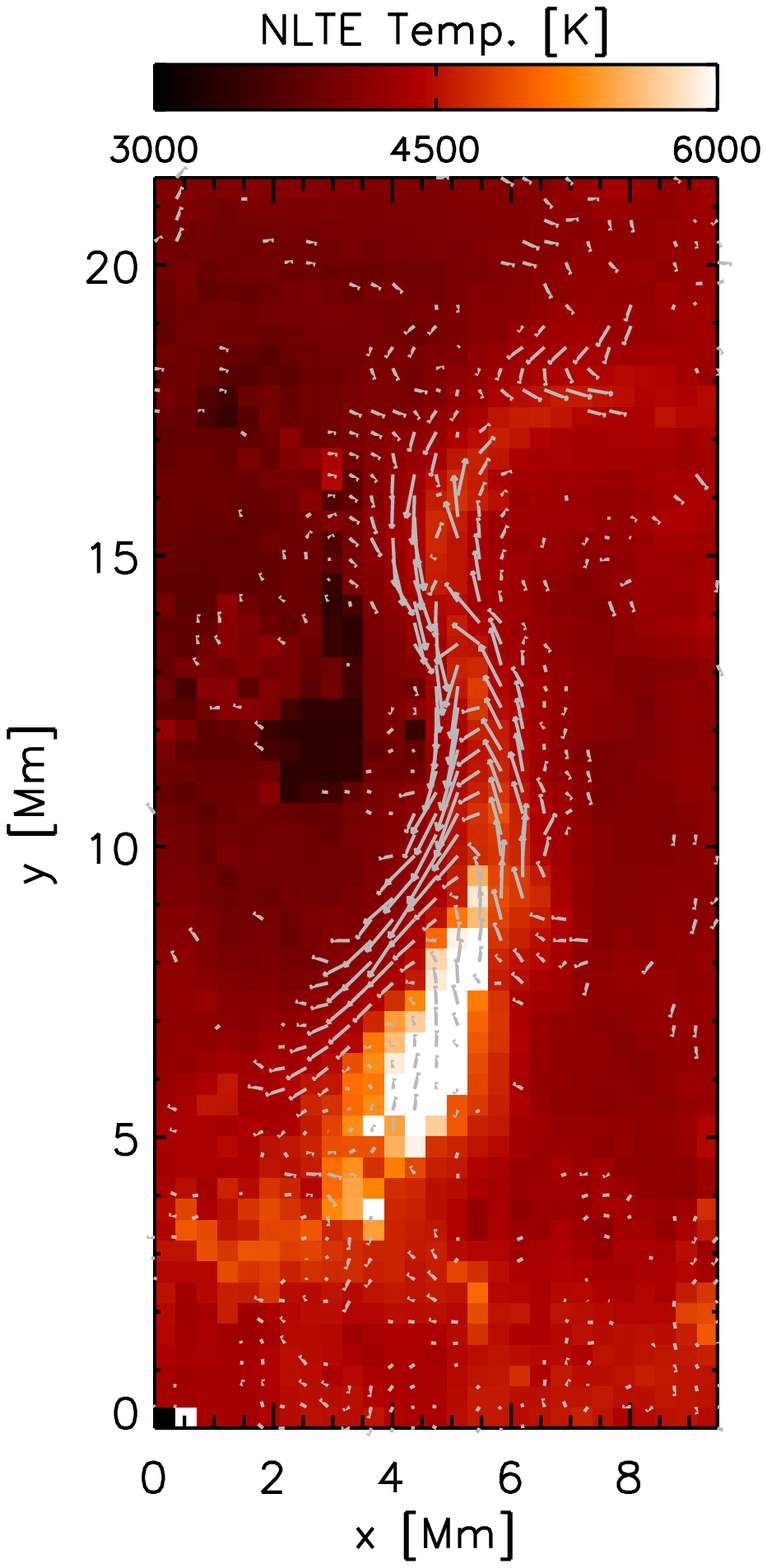}
\hspace{-165pt}
\includegraphics[angle=0,width = 1.15\columnwidth]{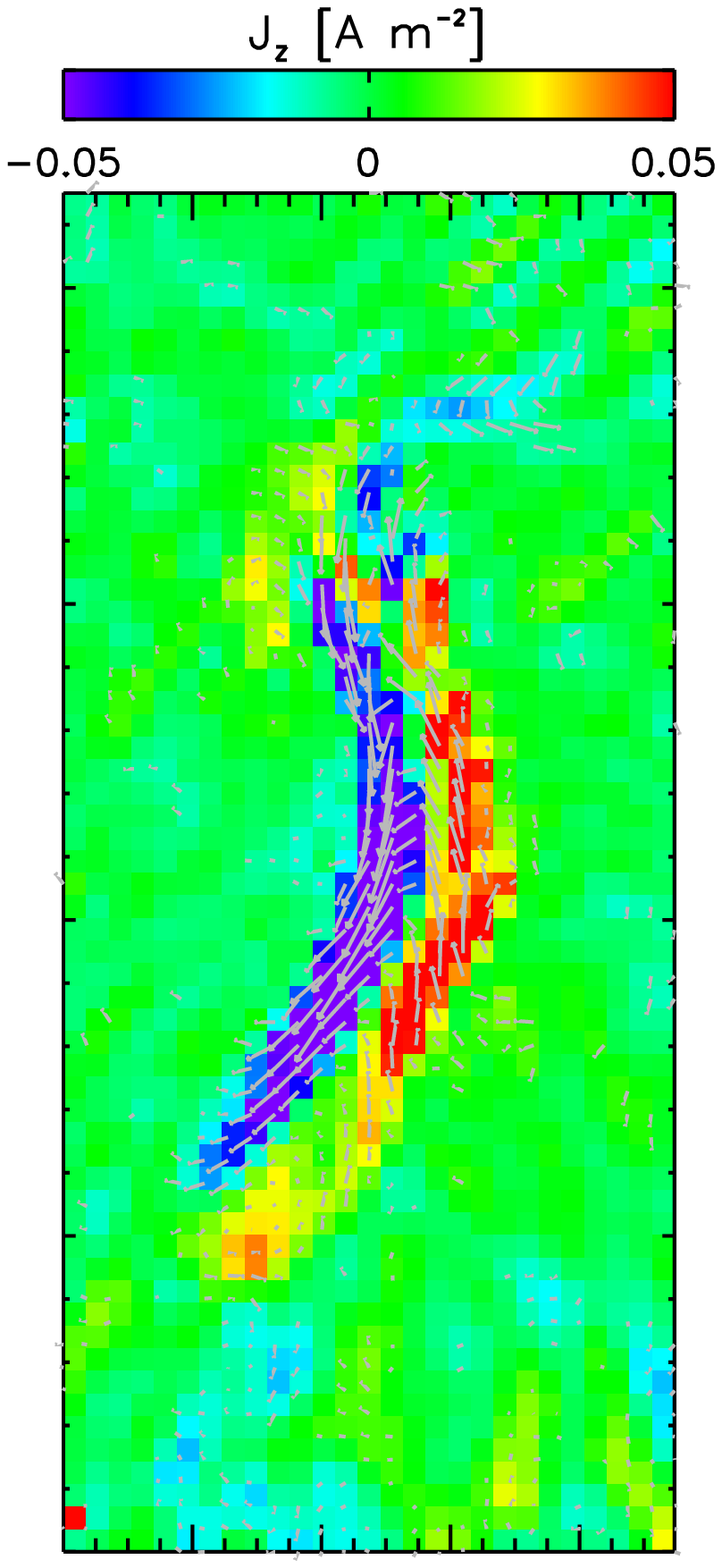}
}
\vspace{-132pt}
\caption{Distribution of electric currents around the temperature enhancements in the LB.
Left: Horizontal current vectors overlaid on the NLTE temperature map 
at 0.36\,Mm for the FOV depicted with the white rectangle in Fig.~\ref{fig02} 
(panel 1a). Arrows have been drawn for every pixel where $|J_z|$ is greater 
than 0.01\,A\,m$^{-2}$. Right: Horizontal current vectors overlaid on $J_z$ at 0\,Mm.} 
\label{fig03}
\end{figure}

\begin{figure*}[btp]
\hspace{50pt}
\begin{minipage}{0.62\textwidth}
\centerline{
\includegraphics[angle=0,width=0.85\textwidth]{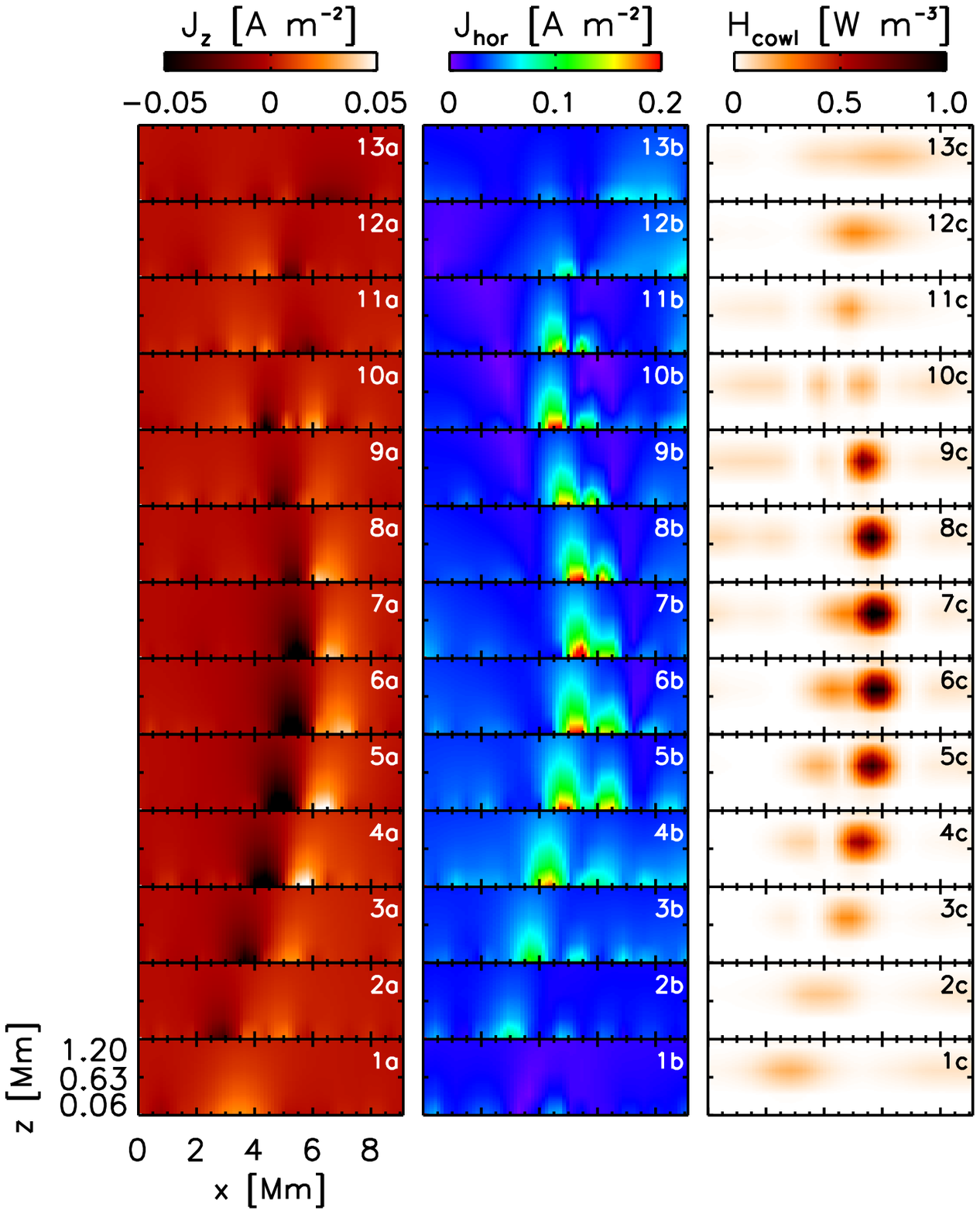}
\hspace{-102pt}
\includegraphics[angle=0,width=0.85\textwidth]{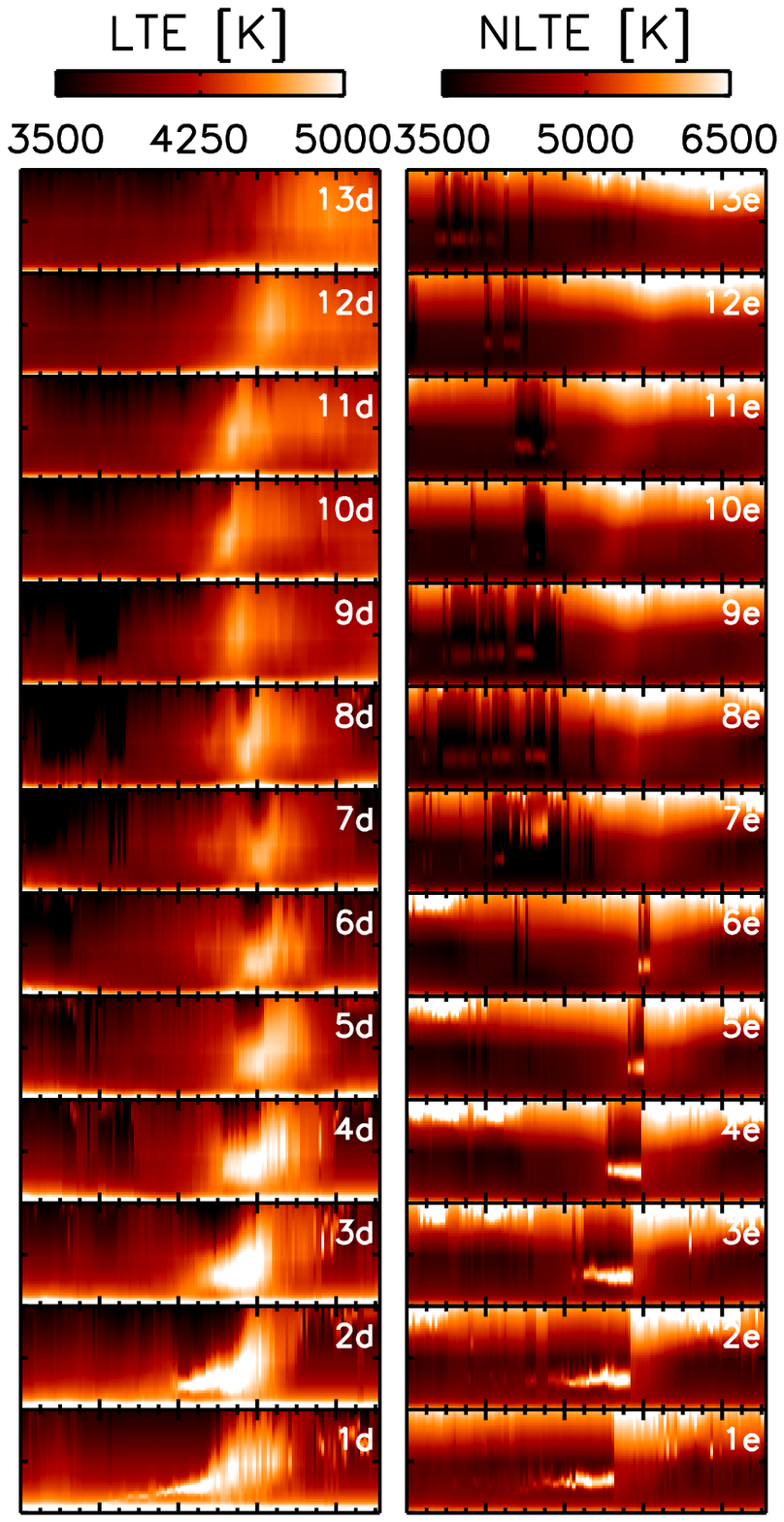}
}
\end{minipage}
\hspace{-60pt}
\begin{minipage}{0.37\textwidth}
\vspace{-105pt}
\centerline{
\includegraphics[angle=0,width = 0.92\textwidth]{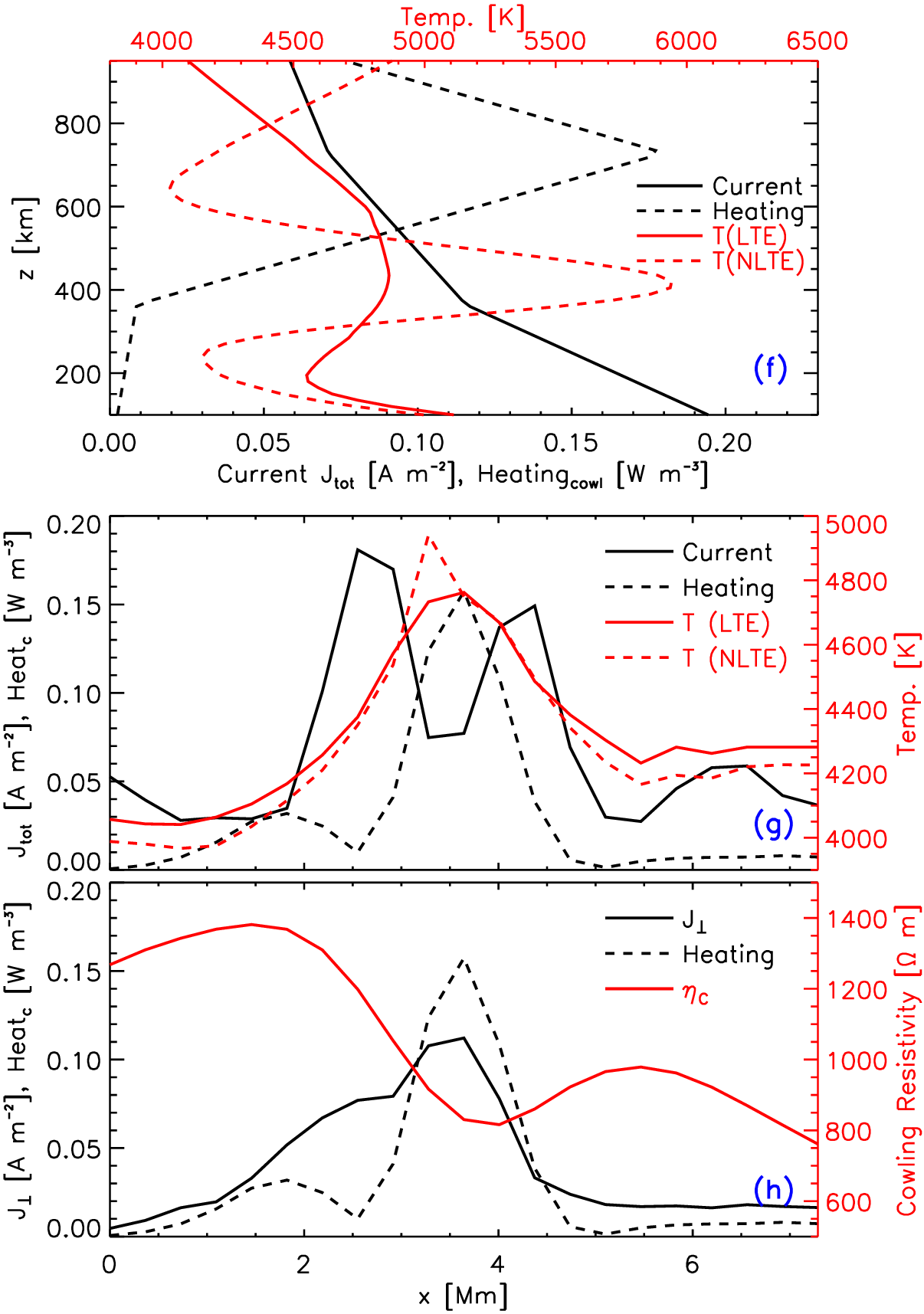}
}
\end{minipage}
\vfill
\vspace{-120pt}
\caption{Vertical variation in physical parameters at different positions on the spine of the LB.
Columns a--e: 2D slices of the currents $J_z$ and $J_{hor}$, Joule heating 
by Cowling resistivity, LTE temperature $T$, and NLTE temperature $T$ (left to right) 
for the horizontal cuts across the LB (panel 2b in Fig.~\ref{fig02}).
Panel f : Vertical distribution of parameters at the center of the LB at the white 
horizontal cut shown in Fig.~\ref{fig02}b. The heating term has been scaled down 
by a factor of five. Panel g: Horizontal variation in parameters along the same cut 
after temperature, current, and heating were averaged in height. The heating term 
has been reduced by a factor of two. Panel h: Same as above but for $J_{\bot}$, 
$\eta_C$, and the Joule heating. $J_{\bot}$ has been scaled up by a factor of three.}
\label{fig04}
\end{figure*}

\begin{figure*}[!ht]
\vspace{10pt}
\centerline{
\includegraphics[angle=0,width = 0.47\textwidth]{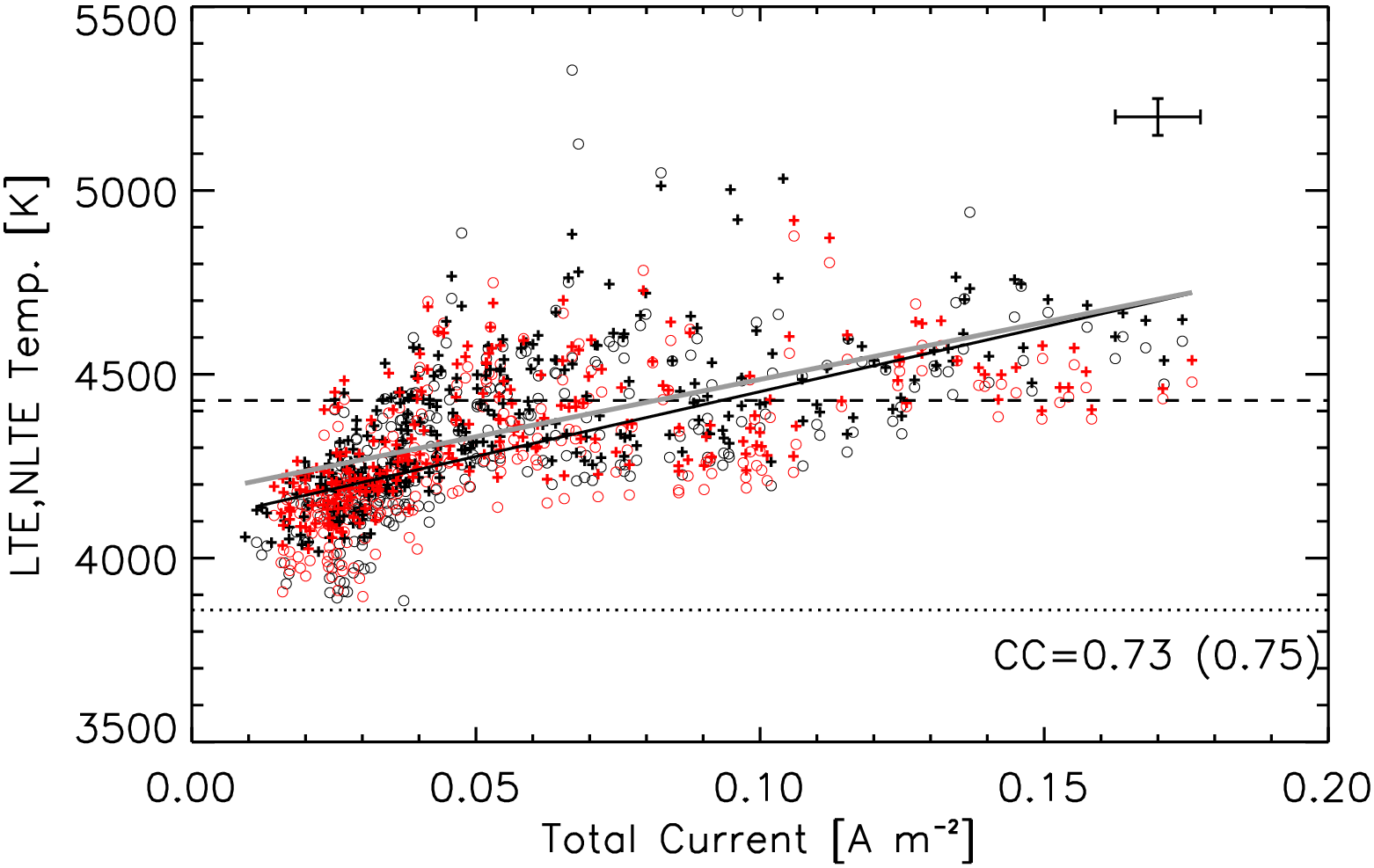}
\hspace{20pt}
\includegraphics[angle=0,width = 0.47\textwidth]{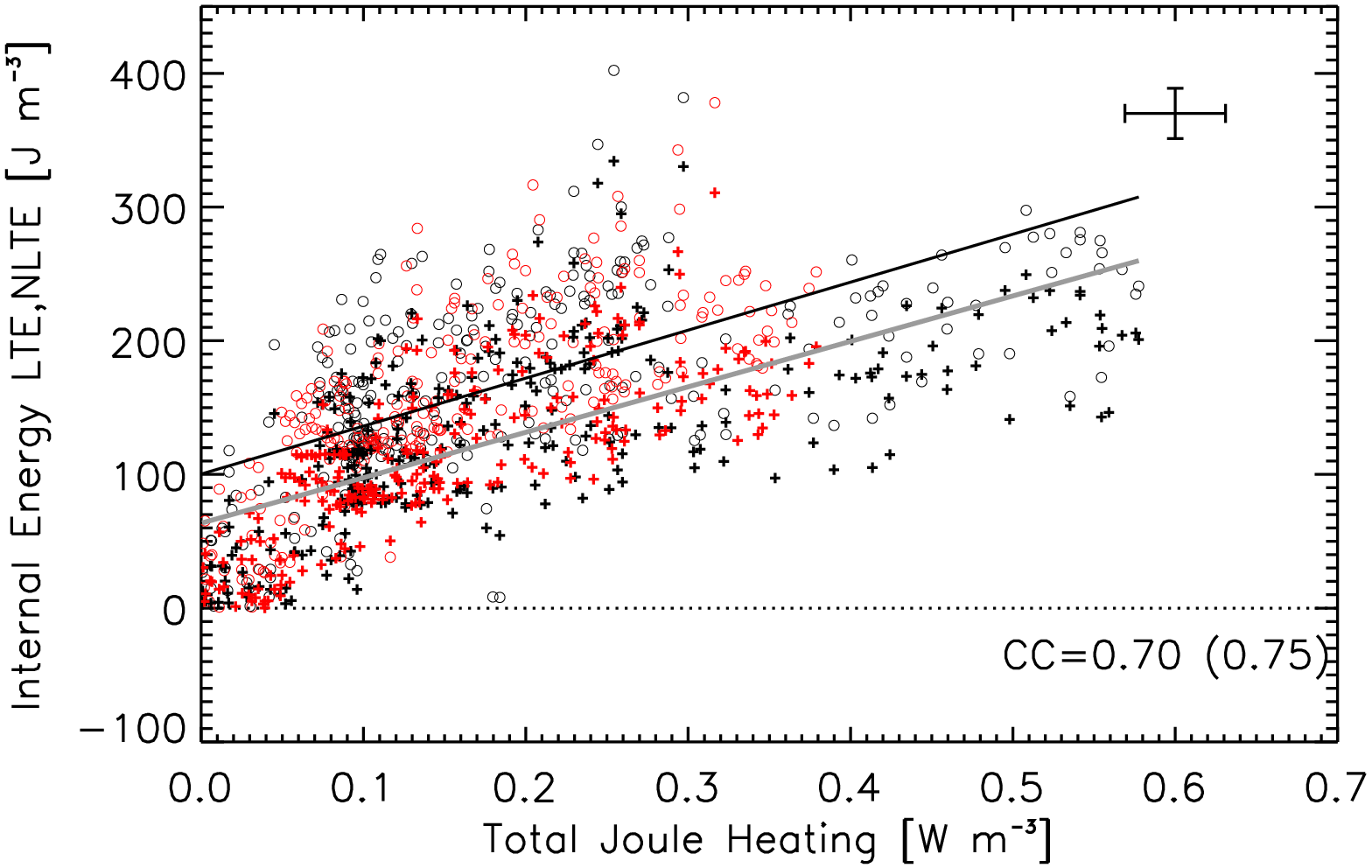}
}
\vspace{-175pt}
\caption{Correlation between currents and temperature enhancements as well as 
Joule heating and increase in internal energy in the LB.
Left:Scatter plots of $|\bf{J}|$ and temperature. The pluses and open circles 
correspond to the LTE and NLTE case, respectively. The red and black colors correspond 
to the data set at 20:48 UT and 21:00 UT, respectively. The solid black line and the grey line
are a linear fit to the scatter in the NLTE and LTE case, respectively. The Spearman correlation coefficient 
for the NLTE case is shown in the lower right corner, while the LTE value is enclosed in 
parentheses. The dotted and dashed lines correspond to the average umbral and quiet Sun 
temperature, respectively.  Right: Scatter plots of Joule heating and internal energy. 
The mean error associated with the different parameters in the LB is shown in the top 
right corner of each panel.} 
\label{fig05}
\end{figure*}

\section{Data analysis}
\label{secana}
To infer the magnetic connectivity, we used the NFFF extrapolation technique 
\citep{hu&dasgupta2008soph,hu+2008apj,hu+2010jastp}, where the magnetic field is 
described by the double-curl Beltrami equation derived from a variational principle 
of the minimum energy dissipation rate \citep{bhattacharyya+2007soph}. This method 
is well suited to the high plasma-$\beta$ photospheric boundary \citep{2001SoPh..203...71G} 
and has been successfully used in many recent studies 
\citep{nayak+2019apj,liu+2020ApJ,yalim+2020ApJ,prasad+2020ApJ}. The extrapolation 
provides the magnetic field vector $\vec{B} (x,y,z)$, from which we derived the total 
($|{\bf J}|$), vertical ($J_z$), and horizontal ($J_{\rm hor}$) current densities. 
The mean values of $B_x$, $B_y$, and $B_z$ in the LB and their errors, as derived from 
the HMI SHARP maps, were 525$\pm$45\,G, 890$\pm$50\,G, and 850$\pm$55\,G, respectively. 
The uncertainty in the current density was determined by using the deviation between 
the extrapolated and observed horizontal magnetic field, which yielded a mean error of 
1.6$\times$10$^{-2}$\,A\,m$^{-2}$, 1.5$\times$10$^{-2}$\,A\,m$^{-2}$, and 
5.0$\times$10$^{-3}$\,A\,m$^{-2}$ in $J_x$, $J_y$, and $J_z$, respectively, 
within the LB. These values translate into relative errors of 33\%, 27\%, and 
16\%, respectively, and pertain to the HMI sampling of 0\farcs5.
The frictional Joule heating profile was calculated from the dissipation of currents 
perpendicular to the magnetic field through the Cowling resistivity 
$\eta_{\textrm{\tiny{C}}}$ ~\citep{yalim+2020ApJ}, where $\eta_{\textrm{\tiny{C}}}$ 
is a function of the magnetic field $\mathbf{B}$, plasma bulk density $\rho$, 
and temperature $T$ as well as the ion and electron number densities $n_i$ and $n_e$ 
in the chromosphere. All values of physical quantities, apart from {\textbf{B}}, were 
taken from the tabulated data of the Maltby-M umbral core model~\citep{1986ApJ...306..284M}.
We discarded the Coulomb resistivity because it is two to three orders of magnitude 
smaller than $\eta_{\textrm{\tiny{C}}}$ \citep[see Fig. 3 of][]{yalim+2020ApJ} 
and only retained the Cowling heating: $\eta_{\textrm{\tiny{C}}}J^2_{\bot}$, 
where $J_{\bot}$ is the component of the current perpendicular to the magnetic field. 
The relative error in the Joule heating was about 34\%, corresponding to about 
0.06\,W\,m$^{-3}$ in the LB.

The IBIS spectra were inverted with both the local thermodynamic equilibrium (LTE) 
and non-LTE (NLTE) version of the CAlcium Inversion based on a Spectral ARchive 
\citep[CAISAR;][]{beck+etal2015,beck+etal2019a} code. Individual spectra are inverted 
on a pixel-by-pixel basis. The model atmospheres in the archives are in hydrostatic 
equilibrium. The resulting temperature stratifications were converted from optical 
depth to geometrical height based on the Harvard Smithsonian reference atmosphere 
\citep{1971SoPh...18..347G} without considering lateral pressure equilibrium. 
The temperature maps were then spatially de-projected to the 
local reference frame to match the HMI SHARP magnetic field in $x$ and $y$.
The LTE and NLTE temperature excess, estimated with respect to the mean umbral 
temperature, were subsequently converted to the increase in internal energy as 
in Eq.\,1 of \cite{2013A&A...553A..73B}, see also \cite{2015A&A...582A.104R}, their Sect. 4.7.
For a partially ionized gas, this value serves as the lower limit 
since the product of the ionization fraction and the ionization potential per unit 
mass is an additional term in the internal energy. The error in temperature is 
20--100\,K for $-5$\,$<$\,log\,$\tau$\,$<$\,$-2$, with nearly identical temperatures in LTE and NLTE 
for $\log\tau>-2$ (LBC20, Sect. 3.3).

\section{Results}
\label{res}
\subsection{Global topology of NOAA AR 12002}
\label{topology}
Figure~\ref{fig01}b shows the sunspot in the chromospheric \ion{Ca}{II} spectral line, with 
a strong brightness enhancement all along the LB. The LB formed as a result of large-scale 
magnetic flux emergence in the sunspot over a duration of 13\,hr, which was seen as strong 
blueshifts of about 1\,km\,s$^{-1}$ all along the body of the LB (LBC20). The LB comprised
weak and highly inclined magnetic fields, with a mean field strength and inclination of 
about 1.5\,kG and 55$^\circ$, respectively, while in the southern part of the structure 
the values are about 1.2\,kG and 85$^\circ$, respectively. The LB has a pronounced curvature 
that connects the southern and northwestern penumbral sections (panel c). Figure~\ref{fig01}d 
shows the coronal magnetic field from the NFFF extrapolation overlaid on a composite of the 
Atmospheric Imaging Assembly \citep[AIA,][]{2012SoPh..275...17L}~171\,\AA\,image and $B_z$. 
Prominent loops connect the sunspot to the opposite polarity 
to the east, while the field lines in the western part of the sunspot are mostly open. No 
loops were found to follow the LB structure, but a few field lines directly 
neighboring the LB showed a small dent away from the LB for $z<1$\,Mm.

\subsection{Electric current density and temperature}
\label{currents}
Figure~\ref{fig02} shows that both the current density and temperature are larger over the 
LB than anywhere else in the sunspot at 21:00 UT. At the bottom boundary the total current 
$|{\bf J}|$ reaches a maximum value of 0.3\,A\,m$^{-2}$ with a mean value of 0.11\,A\,m$^{-2}$. 
The total current is dominated by $J_{hor}$, which is three times stronger than $J_z$.  
At z=0.36\,Mm, the maximum values of $J_z$ and $J_{hor}$ reduce to 0.05\,A\,m$^{-2}$ and 
0.13\,A\,m$^{-2}$, respectively.  
At the same height, the entire length of the LB is hotter than the surroundings, with 
LTE and NLTE temperatures of 4700\,K and 4900\,K, respectively. At the 
southern end of the LB, the temperature reaches 5270\,K and 6775\,K in  LTE and NLTE, respectively. 
The LB is nearly 600--800\,K hotter than the neighboring umbra. Panels 3 and 4 of Fig.~\ref{fig02} 
show that the components of the currents parallel ($J_{||}$) and perpendicular ($J_{\bot}$) to 
the magnetic field are about 0.04\,A\,m$^{-2}$ in the LB at a height of 0.36\,Mm and reduce 
by about 60\% at 0.72\,Mm. The $\eta_{\textrm{\tiny{C}}}$ 
increases by a factor of five from 0.36\,Mm to 0.72\,Mm, although it is weaker in 
the LB relative to the umbra, by a factor of about 1.7, owing to its $B^2$ dependence on 
the magnetic field strength. The Joule heating due to $\eta_{\textrm{\tiny{C}}}$ 
is about 1\,W\,m$^{-3}$ at 0.72\,Mm.

Figure~\ref{fig03} shows that the temperature 
enhancement along the spine of the LB is flanked by intense currents that have opposite 
signs in $J_z$. The $J_{\rm hor}$, depicted with arrows, is parallel to the LB spine, with 
opposite directions on either side. Nearly identical properties were found for the second 
data set at 20:48 UT.

Panels a--e of Fig.~\ref{fig04} show 2D $x$--$z$ plots of different physical parameters 
along the cuts across the LB indicated in panel 2b of Fig.~\ref{fig02}. The bottom and top 
panels (i.e., 1a and 13a) correspond to the cut at the southern and northern end, respectively.
Both current and temperature 
enhancements follow the location of the spine of the LB along its extent. As already 
seen in Figs.~\ref{fig02} and \ref{fig03}, the LB is hottest at its southern end (panels 1--4), 
with temperature decreasing towards its northern end. The temperature enhancement is seen 
in both LTE and NLTE, although in NLTE the enhancement is much hotter and narrower, confined 
to a height between 0.2 and 0.6\,Mm (panels 1d--5d and 1e--5e in Fig.~\ref{fig04}). The 
Joule heating (panels c) also exhibits the same spatial variation along the spine of the LB. 
However, the heating is confined to a narrow height range at about 0.7\,Mm (panels 3c--9c)
because of the vertical stratification of $\eta_{\textrm{\tiny{C}}}$.

The height extent of the different quantities is better seen in Fig.~\ref{fig04}f, which 
shows the stratification of $T$, $|\bf{J}|$, and heating for cut No.~5. The NLTE temperature 
has a sharp peak of up to 2000\,K of 0.4\,Mm width at a height of 0.42\,Mm, while the LTE 
temperature shows a broader, lower enhancement of 800\,K over a larger height range. 
The current drops monotonically with height, while the heating shows a sharp peak, similar 
to the NLTE temperature but at a slightly higher altitude of 0.72\,Mm.  

Figure~\ref{fig04}g, with height-averaged $T$, $|\bf{J}|$, and heating values for 
the same cut, confirms once more that the largest currents flank the LB spine, while the 
temperature enhancements are co-spatial with the Joule heating at the center of the LB. 
The temperature, current, and heating were averaged in height between 0.18--0.6\,Mm, 
0--0.36\,Mm, and 0--0.72\,Mm, respectively. Figure~\ref{fig04}h shows the heating to be 
highest at the LB as $J_{\bot}$ has a maximum value due to the large field inclination in 
the LB even if $\eta_{\textrm{\tiny{C}}}$ decreases slightly. The $J_{\bot}^2$ 
dependence of the Joule heating renders it a factor of six higher in 
the LB than the adjacent umbra. Similar currents and heating 
were already present at 20:36\,UT, about 10\,min prior to the DST observations.

\subsection{Ohmic dissipation in the LB}
\label{ohm}
While Figs.~\ref{fig02} and \ref{fig03} demonstrate the spatial association of the 
currents and Joule heating with temperature, Fig.~\ref{fig05} shows the scatter plots 
of currents, heating, and temperature. The scatter plots were constructed by taking a 13-pixel-wide 
horizontal cut, centered on the location of maximum temperature on the LB spine, and then 
averaging temperature, the total current, and the Joule heating over height as described 
in the previous section. To cover the spatial extent of the LB, we moved the location of 
these cuts in steps of one pixel from south to north. The left panel of Fig.~\ref{fig05} 
reveals that the total current $|{\bf J}|$ is strongly positively correlated with the 
temperature, exhibiting a Spearman coefficient of 0.75 and 0.73 for the LTE and NLTE case, 
respectively. The scatter plots between the increase in internal energy over the neighboring 
umbra and the Joule heating (right panel of Fig.~\ref{fig05}) also exhibit a high correlation, 
with values of 0.75 and 0.7 for the LTE and NLTE case, respectively.
  
The ratio of the increase in internal energy and the Joule heating provides the timescale 
over which the currents must dissipate in order to produce the observed temperature 
enhancements. The dissipative timescale is about 10\,min, with both the LTE and NLTE 
cases having a nearly identical slope in the linear fit. This is consistent with the 
lifetime of the currents that are permanently present from 20:36\,UT to 21:00\,UT. 
The excess emission in the \ion{Ca}{II}\,IR line core in the LB relative to the quiet 
Sun amounted to only about 6$\times 10^{-3}$\,W\,m$^{-3}$ at a characteristic heating 
rate of 0.2\,W\,m$^{-3}$. Radiative losses should thus have no strong impact on 
the timescale.

\section{Discussion}
\label{discuss}
We find that the LB is hotter than the rest of the sunspot, both spatially and vertically. The 
temperature enhancement is driven by intense electric currents, with the horizontal component 
dominating the vertical current. The maximum value of $J_z$ at $z=0$\,Mm is about 0.1\,A\,m$^{-2}$, 
which is consistent with  \cite{2015ApJ...811..137T}. These values are about a factor of two smaller 
than those reported by\cite{2006A&A...453.1079J}, \cite{2006A&A...449.1169B}, and \cite{2009ApJ...696L..66S},
which could be due to the HMI spatial resolution in our case being insufficient 
to resolve current sheets. The true currents are expected to be larger than the values derived.
The $|\bf{J}|$ had maximum and mean values of 0.2 and 0.14\,A\,m$^{-2}$, respectively, in the penumbra 
in \cite{2010ApJ...721L..58P}. The large 
values of 0.3\,A\,m$^{-2}$ estimated in the LB arise from the emergence of a relatively weak, 
nearly horizontal magnetic structure in the sunspot over a duration of at least 13~hrs (LBC20). 
As a consequence of the sustained emergence of horizontal magnetic fields, one would expect the 
corresponding temperature enhancements to likely be present over a similar timescale. The value 
of the current density in the rest of the sunspot is consistent 
with the results of \cite{2005ApJ...633L..57S} and \cite{2010ApJ...721L..58P}. 
Flux emergence thus seems to be a good indicator for locations, with a strong correlation between 
currents and temperature enhancements. While the strong correlation between currents
and increased temperature is seen for 
most of the LB spine, the strongest temperature enhancements at the southern end of the 
LB are associated with weaker currents (Figs.~\ref{fig02} and ~\ref{fig03}). This suggests that 
the heating mechanism in the LB could be more intricate than just a strongly localized current 
dissipation with an instant temperature rise. The complex structure of the LB and its emerging 
magnetic flux (which is not fully resolved in the HMI data), the magnetic field extrapolation, 
directed mass flows, heat conduction, and possible time lags between heating and temperature 
could lead to spatio-temporal displacements between currents and enhanced temperatures.   

The LTE and NLTE temperature maps reveal that the thermal 
enhancements are confined to a comparably narrow slab in the vertical direction, especially 
in the NLTE inversion. The height resolution of the spectra is limited to a few hundred kilometers by the 
thermal broadening \citep{2012A&A...544A..46B,2013A&A...553A..73B}, which smears out current 
sheets with heights of only a few kilometers. 
\cite{2015ApJ...811..138T} reported a shift in the $\log\tau=1$ layer by a few hundred kilometers in height 
between an LB and the umbra. This is still within a single pixel in height in the extrapolation 
results that employ the HMI sampling of about 360\,km in the vertical axis as well. Our approach of 
averaging over height reduces the impact of all effects related to the specific geometrical 
heights attributed to the different quantities.

The currents are strongest at the bottom boundary
and decrease monotonically with height. 
However, the Joule heating term, which is dominated by the Cowling 
resistivity, tends to have strong peaks at two heights, 0.72\,Mm and 1.8\,Mm 
\citep{yalim+2020ApJ}, and only the lower one is inside the \ion{Ca}{II} IR formation 
height. The height of the temperature enhancements is 
close to the lower peak, where Joule heating is prevalent (Fig~\ref{fig04}), well within the 
uncertainties associated with the geometrical height scale provided by the spectral
inversions or the tabulated Maltby umbral model. We note that 
in the cool Maltby model the simplified calculation with a hydrogen-only 
atmosphere and an approximate estimate of the ionization degree in the derivation of 
$\eta_{\textrm{\tiny{C}}}$ \citep{yalim+2020ApJ} can cause additional errors in the 
strength and location in height of the Joule heating.    

We find a strong, overall spatial association and a high correlation between  
Ohmic dissipation and the increase in internal energy, which would also be important 
in the context of how small-scale flux emergence \citep{2015A&A...584A...1L} or 
magnetic inhomogeneities in sunspots \citep{2009ApJ...704L..29L,2015AdSpR..56.2305L} 
heat the chromosphere. Along with the estimated dissipative timescale that matches 
the duration of the increased temperatures, the above results are direct evidence 
of chromospheric heating by Ohmic dissipation, which, to the best of our knowledge, 
has not been demonstrated in previous investigations. This process could thus be 
a valid heating source not only in the solar chromosphere but also in lab plasmas 
and other magnetically active stars wherever appropriate magnetic field strengths 
and densities exist.  

\section{Conclusions}
\label{conclu}
The large chromospheric temperature enhancement in the sunspot at the location of 
the LB arises from strong electric currents that are caused by the magnetic 
configuration of the LB, where relatively weak and highly inclined magnetic 
fields emerge over a duration of about 13 hr. The temperature excess due to the 
dissipation of the currents is located in the lower chromosphere between 0.4--0.7\,Mm 
and is possibly sustained over the whole passage of flux emergence. The characteristic 
timescale for the heating is about 10\,min. Our study provides direct evidence of 
lower chromospheric heating by the Ohmic dissipation of electric currents in a sunspot.

\begin{acknowledgements}
The Dunn Solar Telescope at Sacramento Peak/NM was 
operated by the National Solar Observatory (NSO). NSO is operated by the Association 
of Universities for Research in Astronomy (AURA), Inc. under cooperative agreement 
with the National Science Foundation (NSF). HMI data are courtesy of NASA/SDO and 
the HMI science team.  They are provided by the Joint Science Operations Center -- Science
Data Processing at Stanford University. IBIS has been designed and constructed by 
the INAF/Osservatorio Astrofisico di Arcetri with contributions from the Universit{\`a} 
di Firenze, the Universit{\`a}di Roma Tor Vergata, and upgraded with further contributions 
from NSO and Queens University Belfast. This work was supported through NSF grant
AGS-1413686. M.S.Y. and A.P. acknowledge partial support from NSF award AGS-2020703. 
M.S.Y. also acknowledges partial support from NASA LWS grant 80NSSC19K0075 and the 
NSF EPSCoR RII-Track-1 Cooperative Agreement OIA-1655280. We would like
to thank the referee for reviewing our article and for providing insightful comments.
\end{acknowledgements}

\bibliographystyle{aa}
\bibliography{louis_ref}

\begin{thebibliography}{55}
\expandafter\ifx\csname natexlab\endcsname\relax\def\natexlab#1{#1}\fi

\bibitem[{{Athay} \& {Holzer}(1982)}]{1982ApJ...255..743A}
{Athay}, R.~G. \& {Holzer}, T.~E. 1982, \apj, 255, 743

\bibitem[{{Bahauddin} {et~al.}(2021){Bahauddin}, {Bradshaw}, \&
  {Winebarger}}]{2021NatAs...5..237B}
{Bahauddin}, S.~M., {Bradshaw}, S.~J., \& {Winebarger}, A.~R. 2021, Nature
  Astronomy, 5, 237

\bibitem[{{Balthasar}(2006)}]{2006A&A...449.1169B}
{Balthasar}, H. 2006, \aap, 449, 1169

\bibitem[{{Balthasar} {et~al.}(2014){Balthasar}, {Beck}, {Louis}, {Verma}, \&
  {Denker}}]{2014A&A...562L...6B}
{Balthasar}, H., {Beck}, C., {Louis}, R.~E., {Verma}, M., \& {Denker}, C. 2014,
  \aap, 562, L6

\bibitem[{{Beck} {et~al.}(2015){Beck}, {Choudhary}, {Rezaei}, \&
  {Louis}}]{beck+etal2015}
{Beck}, C., {Choudhary}, D.~P., {Rezaei}, R., \& {Louis}, R.~E. 2015, \apj,
  798, 100

\bibitem[{{Beck} {et~al.}(2019){Beck}, {Gosain}, \&
  {Kiessner}}]{beck+etal2019a}
{Beck}, C., {Gosain}, S., \& {Kiessner}, C. 2019, \apj, 878, 60

\bibitem[{{Beck} {et~al.}(2009){Beck}, {Khomenko}, {Rezaei}, \&
  {Collados}}]{2009A&A...507..453B}
{Beck}, C., {Khomenko}, E., {Rezaei}, R., \& {Collados}, M. 2009, \aap, 507,
  453

\bibitem[{{Beck} {et~al.}(2012){Beck}, {Rezaei}, \&
  {Puschmann}}]{2012A&A...544A..46B}
{Beck}, C., {Rezaei}, R., \& {Puschmann}, K.~G. 2012, \aap, 544, A46

\bibitem[{{Beck} {et~al.}(2013){Beck}, {Rezaei}, \&
  {Puschmann}}]{2013A&A...553A..73B}
{Beck}, C., {Rezaei}, R., \& {Puschmann}, K.~G. 2013, \aap, 553, A73

\bibitem[{{Beck} {et~al.}(2016){Beck}, {Rezaei}, {Puschmann}, \&
  {Fabbian}}]{2016SoPh..291.2281B}
{Beck}, C., {Rezaei}, R., {Puschmann}, K.~G., \& {Fabbian}, D. 2016, \solphys,
  291, 2281

\bibitem[{{Beckers}(1968)}]{1968SoPh....3..367B}
{Beckers}, J.~M. 1968, \solphys, 3, 367

\bibitem[{{Bhattacharyya} {et~al.}(2007){Bhattacharyya}, {Janaki}, {Dasgupta},
  \& {Zank}}]{bhattacharyya+2007soph}
{Bhattacharyya}, R., {Janaki}, M.~S., {Dasgupta}, B., \& {Zank}, G.~P. 2007,
  \solphys, 240, 63

\bibitem[{{Cavallini}(2006)}]{cavallini2006}
{Cavallini}, F. 2006, \solphys, 236, 415

\bibitem[{{De Pontieu} {et~al.}(2015){De Pontieu}, {McIntosh},
  {Martinez-Sykora}, {Peter}, \& {Pereira}}]{2015ApJ...799L..12D}
{De Pontieu}, B., {McIntosh}, S., {Martinez-Sykora}, J., {Peter}, H., \&
  {Pereira}, T.~M.~D. 2015, \apjl, 799, L12

\bibitem[{{De Pontieu} {et~al.}(2009){De Pontieu}, {McIntosh}, {Hansteen}, \&
  {Schrijver}}]{2009ApJ...701L...1D}
{De Pontieu}, B., {McIntosh}, S.~W., {Hansteen}, V.~H., \& {Schrijver}, C.~J.
  2009, \apjl, 701, L1

\bibitem[{{Gary}(2001)}]{2001SoPh..203...71G}
{Gary}, G.~A. 2001, \solphys, 203, 71

\bibitem[{{Gingerich} {et~al.}(1971){Gingerich}, {Noyes}, {Kalkofen}, \&
  {Cuny}}]{1971SoPh...18..347G}
{Gingerich}, O., {Noyes}, R.~W., {Kalkofen}, W., \& {Cuny}, Y. 1971, \solphys,
  18, 347

\bibitem[{{Grant} {et~al.}(2018){Grant}, {Jess}, {Zaqarashvili}, {Beck},
  {Socas-Navarro}, {Aschwanden}, {Keys}, {Christian}, {Houston}, \&
  {Hewitt}}]{2018NatPh..14..480G}
{Grant}, S. D.~T., {Jess}, D.~B., {Zaqarashvili}, T.~V., {et~al.} 2018, Nature
  Physics, 14, 480

\bibitem[{{Hu} \& {Dasgupta}(2008)}]{hu&dasgupta2008soph}
{Hu}, Q. \& {Dasgupta}, B. 2008, \solphys, 247, 87

\bibitem[{{Hu} {et~al.}(2008){Hu}, {Dasgupta}, {Choudhary}, \&
  {B{\"u}chner}}]{hu+2008apj}
{Hu}, Q., {Dasgupta}, B., {Choudhary}, D.~P., \& {B{\"u}chner}, J. 2008, \apj,
  679, 848

\bibitem[{{Hu} {et~al.}(2010){Hu}, {Dasgupta}, {Derosa}, {B{\"u}chner}, \&
  {Gary}}]{hu+2010jastp}
{Hu}, Q., {Dasgupta}, B., {Derosa}, M.~L., {B{\"u}chner}, J., \& {Gary}, G.~A.
  2010, Journal of Atmospheric and Solar-Terrestrial Physics, 72, 219

\bibitem[{{Jur{\v{c}}{\'a}k} {et~al.}(2006){Jur{\v{c}}{\'a}k}, {Mart{\'\i}nez
  Pillet}, \& {Sobotka}}]{2006A&A...453.1079J}
{Jur{\v{c}}{\'a}k}, J., {Mart{\'\i}nez Pillet}, V., \& {Sobotka}, M. 2006,
  \aap, 453, 1079

\bibitem[{{Kalkofen}(2007)}]{2007ApJ...671.2154K}
{Kalkofen}, W. 2007, \apj, 671, 2154

\bibitem[{{Lemen} {et~al.}(2012){Lemen}, {Title}, {Akin}, {Boerner}, {Chou},
  {Drake}, {Duncan}, {Edwards}, {Friedlaender}, {Heyman}, {Hurlburt}, {Katz},
  {Kushner}, {Levay}, {Lindgren}, {Mathur}, {McFeaters}, {Mitchell}, {Rehse},
  {Schrijver}, {Springer}, {Stern}, {Tarbell}, {Wuelser}, {Wolfson}, {Yanari},
  {Bookbinder}, {Cheimets}, {Caldwell}, {Deluca}, {Gates}, {Golub}, {Park},
  {Podgorski}, {Bush}, {Scherrer}, {Gummin}, {Smith}, {Auker}, {Jerram},
  {Pool}, {Soufli}, {Windt}, {Beardsley}, {Clapp}, {Lang}, \&
  {Waltham}}]{2012SoPh..275...17L}
{Lemen}, J.~R., {Title}, A.~M., {Akin}, D.~J., {et~al.} 2012, \solphys, 275, 17

\bibitem[{{Liu} {et~al.}(2020){Liu}, {Prasad}, {Lee}, \& {Wang}}]{liu+2020ApJ}
{Liu}, C., {Prasad}, A., {Lee}, J., \& {Wang}, H. 2020, \apj, 899, 34

\bibitem[{{Louis}(2015)}]{2015AdSpR..56.2305L}
{Louis}, R.~E. 2015, Advances in Space Research, 56, 2305

\bibitem[{{Louis} {et~al.}(2020){Louis}, {Beck}, \&
  {Choudhary}}]{2020ApJ...905..153L}
{Louis}, R.~E., {Beck}, C., \& {Choudhary}, D.~P. 2020, \apj, 905, 153

\bibitem[{{Louis} {et~al.}(2015){Louis}, {Bellot Rubio}, {de la Cruz
  Rodr{\'\i}guez}, {Socas-Navarro}, \& {Ortiz}}]{2015A&A...584A...1L}
{Louis}, R.~E., {Bellot Rubio}, L.~R., {de la Cruz Rodr{\'\i}guez}, J.,
  {Socas-Navarro}, H., \& {Ortiz}, A. 2015, \aap, 584, A1

\bibitem[{{Louis} {et~al.}(2009){Louis}, {Bellot Rubio}, {Mathew}, \&
  {Venkatakrishnan}}]{2009ApJ...704L..29L}
{Louis}, R.~E., {Bellot Rubio}, L.~R., {Mathew}, S.~K., \& {Venkatakrishnan},
  P. 2009, \apjl, 704, L29

\bibitem[{{Maltby} {et~al.}(1986){Maltby}, {Avrett}, {Carlsson},
  {Kjeldseth-Moe}, {Kurucz}, \& {Loeser}}]{1986ApJ...306..284M}
{Maltby}, P., {Avrett}, E.~H., {Carlsson}, M., {et~al.} 1986, \apj, 306, 284

\bibitem[{{Narain} \& {Ulmschneider}(1996)}]{1996SSRv...75..453N}
{Narain}, U. \& {Ulmschneider}, P. 1996, \ssr, 75, 453

\bibitem[{{Nayak} {et~al.}(2019){Nayak}, {Bhattacharyya}, {Prasad}, {Hu},
  {Kumar}, \& {Joshi}}]{nayak+2019apj}
{Nayak}, S.~S., {Bhattacharyya}, R., {Prasad}, A., {et~al.} 2019, \apj, 875, 10

\bibitem[{{Osterbrock}(1961)}]{1961ApJ...134..347O}
{Osterbrock}, D.~E. 1961, \apj, 134, 347

\bibitem[{{Parker}(1983)}]{1983ApJ...264..635P}
{Parker}, E.~N. 1983, \apj, 264, 635

\bibitem[{{Pesnell} {et~al.}(2012){Pesnell}, {Thompson}, \&
  {Chamberlin}}]{2012SoPh..275....3P}
{Pesnell}, W.~D., {Thompson}, B.~J., \& {Chamberlin}, P.~C. 2012, \solphys,
  275, 3

\bibitem[{{Pneuman} \& {Kopp}(1978)}]{1978SoPh...57...49P}
{Pneuman}, G.~W. \& {Kopp}, R.~A. 1978, \solphys, 57, 49

\bibitem[{{Prasad} {et~al.}(2020){Prasad}, {Dissauer}, {Hu}, {Bhattacharyya},
  {Veronig}, {Kumar}, \& {Joshi}}]{prasad+2020ApJ}
{Prasad}, A., {Dissauer}, K., {Hu}, Q., {et~al.} 2020, \apj, 903, 129

\bibitem[{{Priest} {et~al.}(2018){Priest}, {Chitta}, \&
  {Syntelis}}]{2018ApJ...862L..24P}
{Priest}, E.~R., {Chitta}, L.~P., \& {Syntelis}, P. 2018, \apjl, 862, L24

\bibitem[{{Puschmann} {et~al.}(2010){Puschmann}, {Ruiz Cobo}, \& {Mart{\'\i}nez
  Pillet}}]{2010ApJ...721L..58P}
{Puschmann}, K.~G., {Ruiz Cobo}, B., \& {Mart{\'\i}nez Pillet}, V. 2010, \apjl,
  721, L58

\bibitem[{{Rezaei} \& {Beck}(2015)}]{2015A&A...582A.104R}
{Rezaei}, R. \& {Beck}, C. 2015, \aap, 582, A104

\bibitem[{{Robustini} {et~al.}(2018){Robustini}, {Leenaarts}, \& {de la Cruz
  Rodr{\'\i}guez}}]{2018A&A...609A..14R}
{Robustini}, C., {Leenaarts}, J., \& {de la Cruz Rodr{\'\i}guez}, J. 2018,
  \aap, 609, A14

\bibitem[{{Sakaue} \& {Shibata}(2020)}]{2020ApJ...900..120S}
{Sakaue}, T. \& {Shibata}, K. 2020, \apj, 900, 120

\bibitem[{{Schou} {et~al.}(2012){Schou}, {Scherrer}, {Bush}, {Wachter},
  {Couvidat}, {Rabello-Soares}, {Bogart}, {Hoeksema}, {Liu}, {Duvall}, {Akin},
  {Allard}, {Miles}, {Rairden}, {Shine}, {Tarbell}, {Title}, {Wolfson},
  {Elmore}, {Norton}, \& {Tomczyk}}]{2012SoPh..275..229S}
{Schou}, J., {Scherrer}, P.~H., {Bush}, R.~I., {et~al.} 2012, \solphys, 275,
  229

\bibitem[{{Shelyag} {et~al.}(2016){Shelyag}, {Khomenko}, {de Vicente}, \&
  {Przybylski}}]{2016ApJ...819L..11S}
{Shelyag}, S., {Khomenko}, E., {de Vicente}, A., \& {Przybylski}, D. 2016,
  \apjl, 819, L11

\bibitem[{{Shimizu} {et~al.}(2009){Shimizu}, {Katsukawa}, {Kubo}, {Lites},
  {Ichimoto}, {Suematsu}, {Tsuneta}, {Nagata}, {Shine}, \&
  {Tarbell}}]{2009ApJ...696L..66S}
{Shimizu}, T., {Katsukawa}, Y., {Kubo}, M., {et~al.} 2009, \apjl, 696, L66

\bibitem[{{Socas-Navarro}(2005)}]{2005ApJ...633L..57S}
{Socas-Navarro}, H. 2005, \apjl, 633, L57

\bibitem[{{Solanki} {et~al.}(2003){Solanki}, {Lagg}, {Woch}, {Krupp}, \&
  {Collados}}]{2003Natur.425..692S}
{Solanki}, S.~K., {Lagg}, A., {Woch}, J., {Krupp}, N., \& {Collados}, M. 2003,
  \nat, 425, 692

\bibitem[{{Stein}(1981)}]{1981ApJ...246..966S}
{Stein}, R.~F. 1981, \apj, 246, 966

\bibitem[{{Syntelis} \& {Priest}(2020)}]{2020ApJ...891...52S}
{Syntelis}, P. \& {Priest}, E.~R. 2020, \apj, 891, 52

\bibitem[{{Toriumi} {et~al.}(2015{\natexlab{a}}){Toriumi}, {Cheung}, \&
  {Katsukawa}}]{2015ApJ...811..138T}
{Toriumi}, S., {Cheung}, M. C.~M., \& {Katsukawa}, Y. 2015{\natexlab{a}}, \apj,
  811, 138

\bibitem[{{Toriumi} {et~al.}(2015{\natexlab{b}}){Toriumi}, {Katsukawa}, \&
  {Cheung}}]{2015ApJ...811..137T}
{Toriumi}, S., {Katsukawa}, Y., \& {Cheung}, M. C.~M. 2015{\natexlab{b}}, \apj,
  811, 137

\bibitem[{{Tritschler} {et~al.}(2008){Tritschler}, {Uitenbroek}, \&
  {Reardon}}]{2008ApJ...686L..45T}
{Tritschler}, A., {Uitenbroek}, H., \& {Reardon}, K. 2008, \apjl, 686, L45

\bibitem[{{Ulmschneider} {et~al.}(1978){Ulmschneider}, {Schmitz}, {Kalkofen},
  \& {Bohn}}]{1978A&A....70..487U}
{Ulmschneider}, R., {Schmitz}, F., {Kalkofen}, W., \& {Bohn}, H.~U. 1978, \aap,
  70, 487

\bibitem[{{van Ballegooijen} {et~al.}(2011){van Ballegooijen}, {Asgari-Targhi},
  {Cranmer}, \& {DeLuca}}]{2011ApJ...736....3V}
{van Ballegooijen}, A.~A., {Asgari-Targhi}, M., {Cranmer}, S.~R., \& {DeLuca},
  E.~E. 2011, \apj, 736, 3

\bibitem[{{Yalim} {et~al.}(2020){Yalim}, {Prasad}, {Pogorelov}, {Zank}, \&
  {Hu}}]{yalim+2020ApJ}
{Yalim}, M.~S., {Prasad}, A., {Pogorelov}, N.~V., {Zank}, G.~P., \& {Hu}, Q.
  2020, \apjl, 899, L4

\end{thebibliography}

\end{document}